\RequirePackage{fix-cm} 

\documentclass[conference,comsoc,final,twocolumn]{IEEEtran}

\makeatletter 
\newcommand\semiHuge{\@setfontsize\semiHuge{18.8}{27.38}}
\makeatother

\usepackage[T1]{fontenc}

\usepackage{balance}

\usepackage{cite}

\ifCLASSINFOpdf
  \usepackage[pdftex]{graphicx}
\else
  \usepackage[dvips]{graphicx}
\fi
\usepackage{caption}
\usepackage{amsmath}

\usepackage[cmintegrals]{newtxmath}

\usepackage{algorithmic}

\usepackage{array}

  \usepackage[caption=false,font=normalsize,labelfont=sf,textfont=sf]{subfig}

\usepackage{fixltx2e}

\usepackage{stfloats}
\usepackage{xcolor}

\IEEEoverridecommandlockouts
\begin{document}
%

\title{\semiHuge{Learning to Code: Coded Caching via Deep Reinforcement Learning}}

\author{
    \IEEEauthorblockN{Navid Naderializadeh\\
    Intel Corporation\\
    navid.naderializadeh@intel.com} \and
    \IEEEauthorblockN{Seyed Mohammad Asghari\\
    University of Southern California\\
    asgharip@usc.edu}
}

\maketitle

\begin{abstract}
We consider a system comprising a file library and a network with a server and multiple users equipped with cache memories. The system operates in two phases: a prefetching phase, where users load their caches with parts of contents from the library, and a delivery phase, where users request files from the library and the server needs to send the uncached parts of the requested files to the users. For the case where the users' caches are arbitrarily loaded, we propose an algorithm based on deep reinforcement learning to minimize the delay of delivering requested contents to the users in the delivery phase. Simulation results demonstrate that our proposed deep reinforcement learning agent learns a coded delivery strategy for sending the requests to the users, which slightly outperforms the state-of-the-art performance in terms of delivery delay, while drastically reducing the computational complexity.
\end{abstract}


%
\IEEEpeerreviewmaketitle

\section{Introduction}
It is predicted that the demand for mobile video contents will increase 9-fold from 2017 to 2022, ultimately accounting for around 80\% of the entire mobile data traffic~\cite{cisco}. One of the most promising ways to cope with this significant portion of data traffic would be to use content caching. In fact, caching leverages the \emph{pre-recorded} nature of video contents in order to store parts of the video files in local on-board memories across the network during off-peak hours, which then helps reduce network congestion during peak traffic hours.

The potential gains of caching have given rise to a plethora of literature on characterizing the fundamental limits of caching and devising algorithms to approach those limits. Maddah-Ali and Niesen, in their seminal work~\cite{maddah2014fundamental}, introduced the concept of \emph{coded caching}, and showed that through a careful placement of contents at users' caches, a single server having access to the entire library of files can form coded packets in order to minimize the delivery delay of the requested packets to multiple users. This idea has been extended to many other scenarios as well, such as decentralized caching~\cite{decent}, caching with non-uniform demands~\cite{niesen2016coded}, online caching~\cite{pedarsani2016online}, and hierarchical content delivery networks~\cite{karamchandani2016hierarchical}.

Most of the past work in the coded caching literature has one common limitation: forming the coded packets in the delivery phase requires a very high computational complexity. In fact, as pointed out in~\cite{asghari}, for arbitrary user cache contents, the problem of finding the {\textit{best} set of coded packets} is NP-hard. The authors in~\cite{asghari} have proposed an approximation algorithm, called size-aware coded multicast (SACM), which is guaranteed to achieve a delivery delay that is within $\log(K)$ of the optimal delay, where $K$ is the number of users. However, the SACM approach is still computationally intensive, requiring an \emph{exponential} worst-case complexity, and is practically not applicable to systems involving more than 10 users.

In this paper, we use learning-based techniques in order to enhance the delay of delivering the requested contents to the users in the network, while also reducing the complexity of coded delivery of cached contents to the users. In particular, as in~\cite{maddah2014fundamental,decent,asghari}, we consider a system with a library of multiple files, a single server, and multiple users each equipped with a cache of memory. The system is assumed to operate in two phases:
\begin{itemize}
\item \emph{Prefetching phase}, where users fill up their cache memories using parts of the files from the library.
\item \emph{Delivery phase}, in which each user reveals its request for one of the files in the library, and the server is responsible for delivering (parts of) the requested files which are not cached at the corresponding users.
\end{itemize}

Similar to~\cite{asghari}, we assume that in the prefetching phase, the users cache contents arbitrarily from the library. Note that this is in contrast with~\cite{maddah2014fundamental,decent}, where the end-to-end system is optimized and the cache contents are carefully placed in order to maximize the coding opportunity in the delivery phase. In this scenario, we propose an approach based on deep reinforcement learning (deep RL), where the server is equipped with an actor-critic agent. We define an episode to be the delivery phase for an arbitrary realization of users' cache contents and requests, which terminates after all the requested files are delivered to the users. As in all RL frameworks, the agent interacts with the system by observing states from the environment, taking actions, and receiving rewards during multiple time steps over the course of the episode.

Under a careful design of states, actions, and rewards, we demonstrate that our proposed approach indeed learns to form coded bits, which are then transmitted by the server and consumed by multiple users at the same time. Our RL-based delivery algorithm matches and even slightly outperforms the SACM algorithm~\cite{asghari} in terms of the delivery delay, while reducing its computational complexity in the inference phase from exponential to \emph{polynomial}.

The rest of this paper is organized as follows. In Section~\ref{sec:relwork}, we provide a brief overview of the related work. In Section~\ref{sec:model}, we describe the system model.
In Section~\ref{sec:method}, we explain our proposed RL-based delivery algorithm in detail, including the design of states, actions, and rewards. In Section~\ref{sec:results}, we illustrate the performance of our proposed approach and compare it with multiple baseline algorithms. Finally, we conclude the paper in Section~\ref{sec:conc}.

\section{Related Work}\label{sec:relwork}
\subsection{Coded Caching}
Following the introduction of the coded caching framework in~\cite{maddah2014fundamental}, there has been a multitude of works studying this architecture in different settings. The survey in~\cite{maddah2016coding} provides an overview of these developments in the past years. Notably, the work in~\cite{niesen2015coded} studies the gains of coded caching for delivery of delay-sensitive contents. The authors in~\cite{yu2017exact, wan2016optimality} study the optimality of coded caching in scenarios where the users are limited to uncoded prefetching of contents from the library. Gains of coded caching have been shown to translate to settings with multiple servers~\cite{shariatpanahi2016multi} and nodes with multiple antennas~\cite{shariatpanahi2017multi}. Furthermore, coded caching has also been shown to be beneficial in more complex network deployments, such as device-to-device (D2D) networks~\cite{ji2015wireless, ji2015fundamental}, combination networks~\cite{wan2018caching}, interference channels~\cite{maddah2015cache, naderializadeh2017fundamental, hachem2018degrees, xu2017fundamental}, and cellular networks~\cite{naderializadeh2019cache}.

\subsection{Caching via Machine Learning}
With the success of machine learning (ML), particularly deep learning, in recent years, researchers have introduced ML-based techniques to improve different aspects of systems equipped with caching. In~\cite{chen2018caching}, the authors use probabilistic
latent semantic analysis and the expectation maximization (EM) algorithm to learn individual user preferences based on content popularity, which then helps optimize the caching policy in a D2D setting. In~\cite{ndikumana2018deep}, the authors propose a deep learning approach for caching entertainment contents for self-driving cars using the multi-access edge computing (MEC) architecture. The authors in~\cite{sadeghi2019reinforcement} propose an approach based on dynamic programming and $Q$-learning to adaptively optimize cache contents at the small cell base stations in a 5G cellular network setting and minimize the total cost across files and time. A deep reinforcement learning approach for caching contents at the base station is proposed in~\cite{zhong2018deep} which maximizes the long-term cache hit rate. The authors in~\cite{somuyiwa2018reinforcement} propose a reinforcement learning algorithm to find near-optimal parametrizations for minimizing the long-term average energy spent by a user that proactively caches contents generated in its surrounding environment. Moreover, in~\cite{jiang2019multi}, the authors use multi-agent reinforcement learning to devise optimal caching policies for the user equipment (UE) devices in D2D networks given the history of content demands.

\section{System Model}\label{sec:model}

\begin{figure}[t]
\center
\includegraphics[trim={3.8in 1.8in 4.8in 1.4in},clip, width=0.4\textwidth]{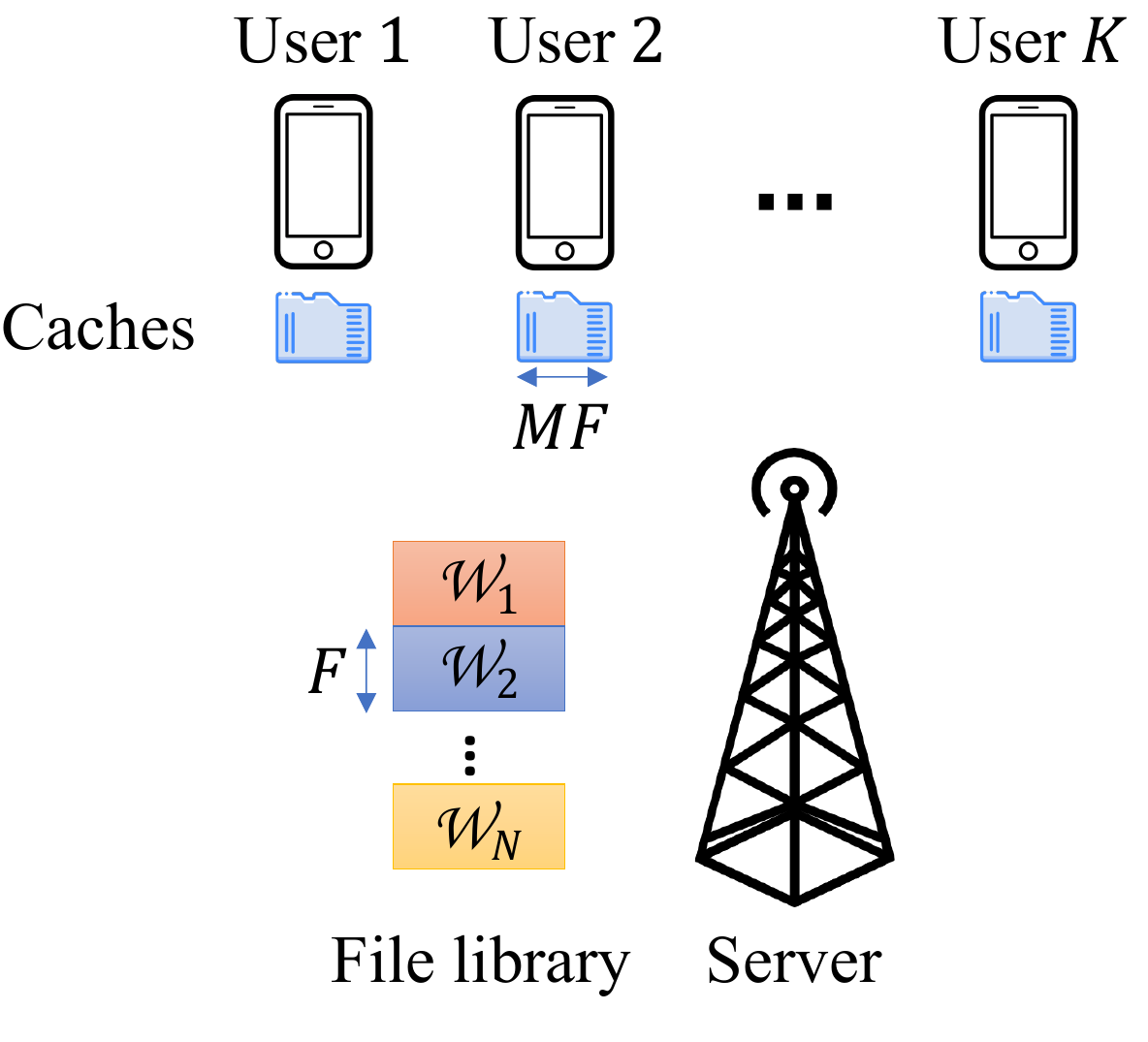}
\caption{System Model of our considered cache-aided network.}
\label{fig:sysModel}
\end{figure}

We consider a network, as illustrated in Figure~\ref{fig:sysModel}, comprising a single server and $K$ users. The server has access to a library of $N$ files $\{\mathcal{W}_1,...,\mathcal{W}_N\}$, each of which consists of $F$ bits. Each user is equipped with a cache of size $M$ files, or equivalently $MF$ bits. The system operates over the following two phases:

\begin{itemize}
\item \emph{Prefetching phase}: In this phase, the caches of all $K$ users across the network are loaded with contents from the library. In particular, each user $i\in\{1,...,K\}$ caches a subset of bits $\mathcal{C}_i$ from the library, whose size is bounded by its cache size; i.e., $|\mathcal{C}_i| \leq MF$. Note that this phase occurs during the off-peak hours, when the network traffic is at a very low level. Moreover, the cache contents of the users may be loaded with or without any knowledge of what contents they will request in the future.

\item \emph{Delivery phase}: In the delivery phase, each user $i\in\{1,...,K\}$ reveals its request for a file $\mathcal{W}_{d_i}, d_i\in\{1,...,N\}$, from the library. Given the cache contents of the users, each user may already have some parts of its desired file in its cache.
Therefore, the server is responsible for delivering the remaining parts of the requested files to the users. We use $\mathcal{D}_i = \mathcal{W}_{d_i} \setminus \mathcal{C}_i$ to represent the subset of file $\mathcal{W}_{d_i}$ that needs to be delivered to user $i$.
\end{itemize}

In order for the server to deliver the bits in $\{\mathcal{D}_i\}_{i=1}^K$ to the corresponding users, we assume that the server broadcasts a sequence of bits $\mathcal{U}=(U_1,U_2,...,U_T)$ to the users. A \emph{normalized delivery delay} of $\frac{T}{F}$ is said to be achievable if there exist a bit sequence $\mathcal{U}=(U_1,U_2,...,U_T)$ and decoding functions $\{f_i(\cdot)\}_{i=1}^K$ such that
\begin{align}\label{eq:decoding}
\mathcal{D}_i = f_i(\mathcal{C}_i, \mathcal{U}), \forall i\in\{1,...,K\}.
\end{align}
Condition \eqref{eq:decoding} essentially implies that each user should be able to decode its requested contents from the library based on its own cache contents and the bits broadcast by the server.

In such a scenario, the ultimate goal is to design the system so as to minimize the normalized delivery delay. Contrary to the coded caching framework, where both the prefetching and delivery phases are jointly optimized to minimize the normalized delivery delay, in the next section, we show how we can leverage deep RL to devise an algorithm for creating a minimal-length bit sequence $\mathcal{U}=(U_1,U_2,...,U_T)$ given \emph{arbitrary} user cache contents in the prefetching phase.

\section{Proposed Algorithm Based on Deep Reinforcement Learning}\label{sec:method}
In this section, we present the details of our proposed approach based on deep reinforcement learning (RL). We equip the server with a deep RL agent, which interacts with the environment (defined as the file library and the users alongside their caches) by observing a state from the environment at each time step, taking an action and receiving a reward, while the environment transitions to the next step. 

We consider an episodic architecture, where each episode is defined as the delivery phase given an arbitrary realization of users' cache contents $\{\mathcal{C}_i\}_{i=1}^K$ and requests $\{d_i\}_{i=1}^K$. The agent interacts with the environment until the delivery phase is complete, in which case the episode is reset and a new set of users' cache contents and requests is realized.

In such a scenario, at each step during an episode, the state, action and reward of the agent are designed as follows:
\begin{itemize}
\item \emph{State}: The state observed by the agent consists of three components:
\begin{itemize}
\item An $NF \times K$ binary \emph{caching} matrix, where each row corresponds to a bit of a file in the library and each column corresponds to a user. Element $(i,j)$ of this matrix is set to 1 if and only if the $i^{th}$ bit in the library is cached by the $j^{th}$ user.

\item An $NF \times K$ binary \emph{requests} matrix, where each row corresponds to a bit in the library and each column corresponds to a user. Element $(i,j)$ of this matrix is set to 1 if and only if the $i^{th}$ bit in the library \emph{still} needs to be delivered to the $j^{th}$ user. Note that once a requested bit is delivered to the requesting user, the corresponding entry in the requests matrix will be set to zero for the rest of the episode.

\item An $NF \times K$ ``\emph{next bit}'' matrix, which is the same as the requests matrix; however, at each step, we only preserve one of the rows of the requests matrix in this matrix corresponding to the next {bit} that needs to be delivered (chosen randomly from all the remaining {bits}) and the rest of the matrix is set to zero. This next bit matrix is included in the state in order to guide the agent to consider sending the selected bit in the coded combination it is going to form at each step.
\end{itemize}
These matrices are concatenated and flattened, implying that the state observed by the agent at each step is a binary vector of length $3NKF$.

\item \emph{Action}: After observing the state at each step, the agent needs to decide which bits the server should send at that step. Therefore, it takes an action in the form of selecting a subset of the bits in the library, which can be represented by a binary vector of length $NF$, equal to the total number of bits in the library. Based on this selection of bits, the server creates a coded bit, which results from the XOR of the selected bits, and broadcasts that coded bit to all users. To reduce the computational complexity (from the exponential complexity of choosing an action from the set of all $2^{NF}$ possible actions), we decouple the $NF$ bits so that for every bit in the library, the agent takes an \emph{individual} action of whether or not to select that bit at that step.

\item \emph{Reward}: After the coded bit is sent by the server, the users will try to decode their desired bits by removing the interference due to the undesired bits using their local cache contents. Denoting the number of successfully delivered bits to all users at the current step by $B$, we have the following two cases for specifying the reward provided to the agent:
\begin{itemize}
\item If the ``next bit'' was delivered successfully, then $\mathsf{reward} = \log_2 B$.
\item Otherwise, $\mathsf{reward} = - 1$.
\end{itemize}
This reward structure encourages the agent to always deliver the ``next bit'' successfully. Moreover, the logarithmic reward regulates the range of the neural network output, which helps improve the training process.
\end{itemize}

\section{Simulation Results}\label{sec:results}
In this section, we analyze the performance of our proposed delivery algorithm based on deep RL and compare it with multiple baseline algorithms. We consider a 2-layer actor-critic agent~\cite{bhatnagar2009natural, degris2012model}, with 32 neurons per layer and a $\tanh$ activation function. We set an upper bound of 100 steps for the episode length (implying that the episode terminates if delivery is not finished within 100 steps), and we use a batch size of 500 steps per training iteration, which includes multiple consecutive episodes as individual trajectories. We add an entropy regularization term with a multiplier of $0.05$ to the loss function in order to encourage the agent to explore more diverse actions to potentially find better subsets of bits to transmit at each time~\cite{williams1991function, mnih2016asynchronous}. We also use an exponential learning rate schedule, starting at a value of $5\times 10^{-3}$ and decaying by $0.9$ every 100 training iterations.

Throughout this section, we assume the worst-case scenario where each user requests a distinct file from the library; i.e., no two users request the same file. Therefore, as the episode begins after the requests are revealed, we only limit the state and action spaces to the $K$ files that have been requested by the users, and discard the $N-K$ remaining files in the library which are irrelevant to that episode. This implies that we can always assume $N=K$, provided that the user requests are distinct.

We compare the performance of our algorithm with the following baseline methods:
\begin{itemize}
\item \emph{Uncoded}: All the bits that are not cached at the requesting users are sent without coding. Hence, this algorithm achieves a normalized delivery delay of $\frac{1}{F} \sum_{i=1}^K |\mathcal{D}_i|$.
\item \emph{Greedy coded multicast (GCM)}~\cite{maddah2014fundamental, decent}: Maddah-Ali and Niesen's delivery algorithm generalized to arbitrary cache placement realizations.
\item \emph{Size-aware coded multicast (SACM)}~\cite{asghari}: The state-of-the-art approximation algorithm that is guaranteed to achieve a normalized delivery delay, which is within a factor $1+\log K$ of the optimum. 
\end{itemize}

\subsection{Normalized Delivery Delay}
\begin{figure}[t]
\center
\includegraphics[width=0.485\textwidth]{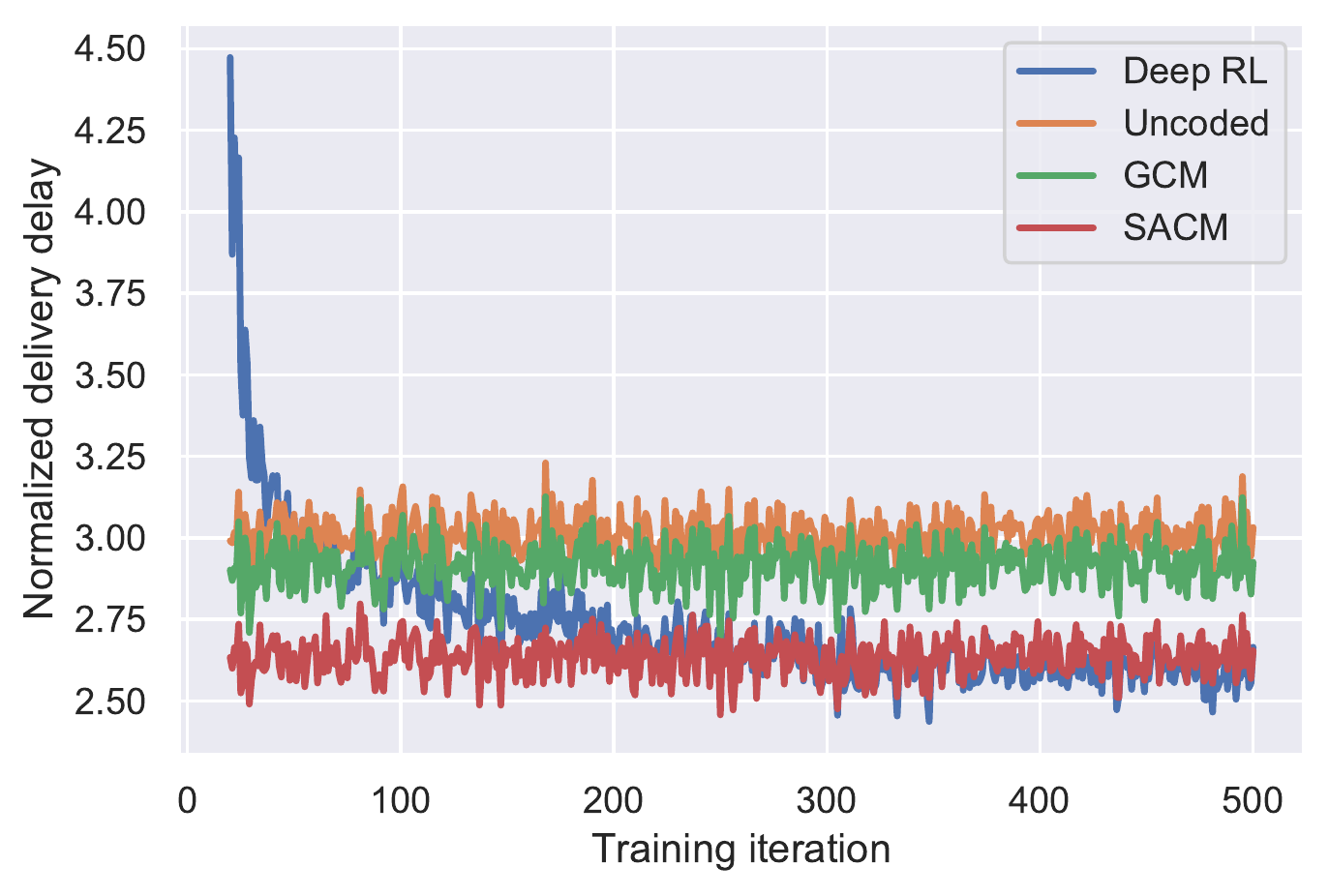}
\caption{Evolution of the normalized delivery delay achieved by our proposed scheme during the training phase and comparison with the baseline algorithms, for a system with $K=4$ users, and a cache size of $M=1$ file at each user.}
\label{fig:delay1}
\end{figure}

Figure \ref{fig:delay1} shows the performance of our proposed deep RL approach over the course of training, where we consider a scenario with $K=4$ users, each capable of caching $M=1$ file.
As the figure depicts, the training procedure can be split into three phases:
\begin{itemize}
\item \emph{Exploration}: At the beginning of training, the agent is initially exploring which bit to send at each time and does not know how to code multiple bits together yet.

\item \emph{Learning to Code}: After approximately 50 training iterations, the agent gains a full understanding of which bits it needs to send to the users. At this point, it starts exploring combinations of different bits to send, and gradually learns how to send more than one bit at each step to maximize its reward, and minimize the normalized delivery delay.

\item \emph{Optimization}: In the final stages of training, after iteration 300, the agent starts to outperform the state-of-the-art SACM approach, and approaches what seems to be the optimal normalized delivery delay in each episode. At this point, the agent fully exploits the coding opportunities to optimize the delivery phase.
\end{itemize}

\begin{figure}[ht]
\center
\includegraphics[width=0.485\textwidth]{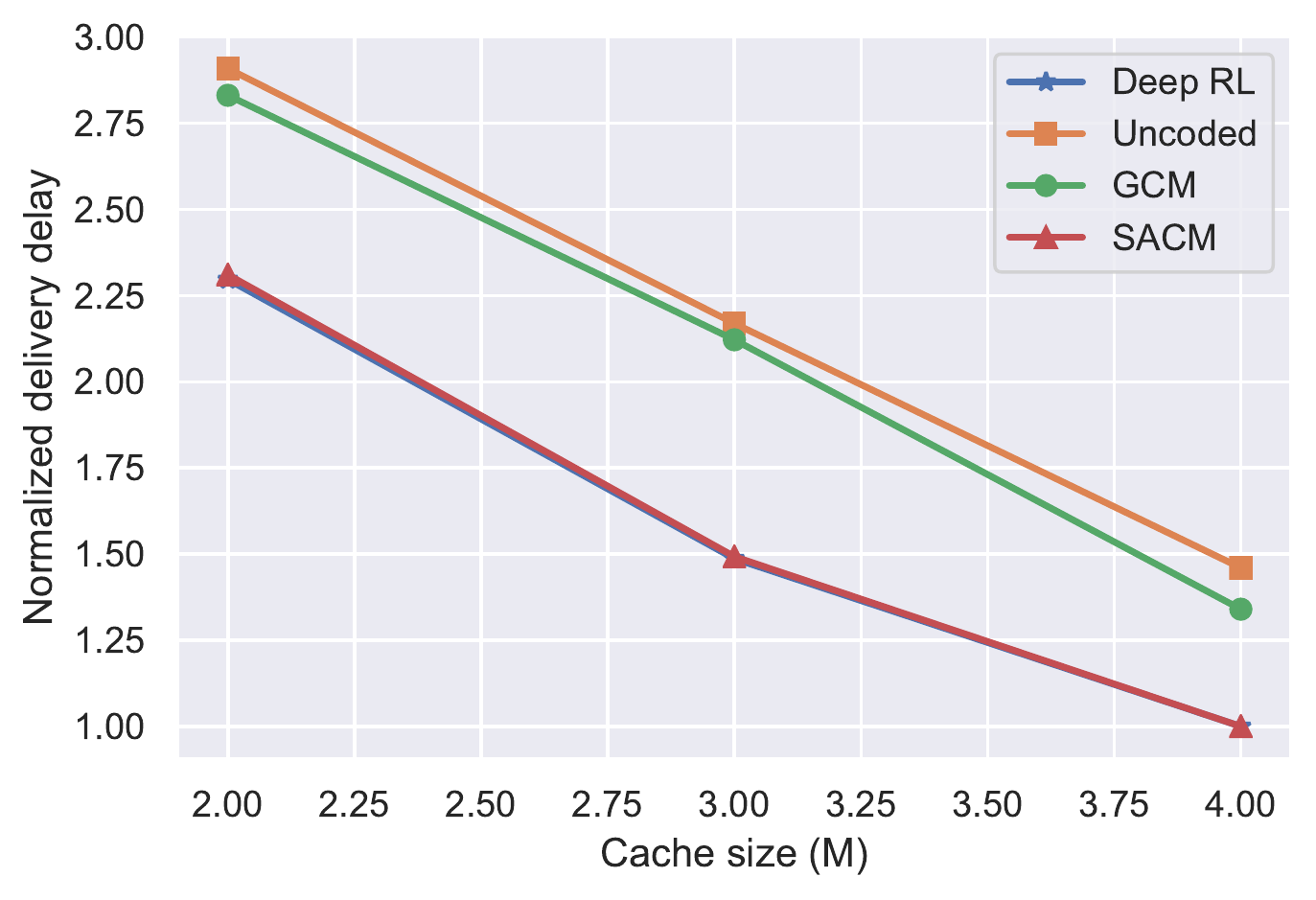}
\caption{Impact of users' cache size on the normalized delivery delay achieved by our proposed scheme and the baseline algorithms for a system with $K=5$ users.}
\label{fig:delay2}
\end{figure}

Moreover, Figure \ref{fig:delay2} illustrates the final performance of our trained actor-critic agent in terms of normalized delivery delay and compares it with the baseline algorithms for a system with $K=5$ users and varying cache size at the users.
As the figure demonstrates, our proposed method can match, and slightly outperform, the performance of the SACM approach, and also achieve significant gains compared to the uncoded and GCM baselines.

\subsection{Computational Complexity}
The main benefit of our proposed approach compared to the SACM approach is in the computational complexity. The SACM approach has a worst-case complexity of $\mathcal{O}(K2^{2K})$, which increases \emph{exponentially} in the number of users~\cite{asghari}. However, our proposed algorithm, once trained, has a worst-case inference complexity of $\mathcal{O}(K^3 F^2)$, which increases \emph{polynomially} in the number of users. This is because the inference complexity per step is $\mathcal{O}(K^2 F) + \mathcal{O}(KF)$, where the first and second terms are due to the state space and action space dimensions, respectively, and there are at most $KF$ steps per episode. Note that the file size $F$ also impacts the complexity of our proposed algorithm. However, depending on the file partitioning method and the cache contents, we may be able to operate on the \emph{chunk} level instead of the bit level, where each chunk contains multiple seconds/minutes of video content~\cite{de2016complexity}.

\begin{figure}[b]
\center
\includegraphics[width=0.485\textwidth]{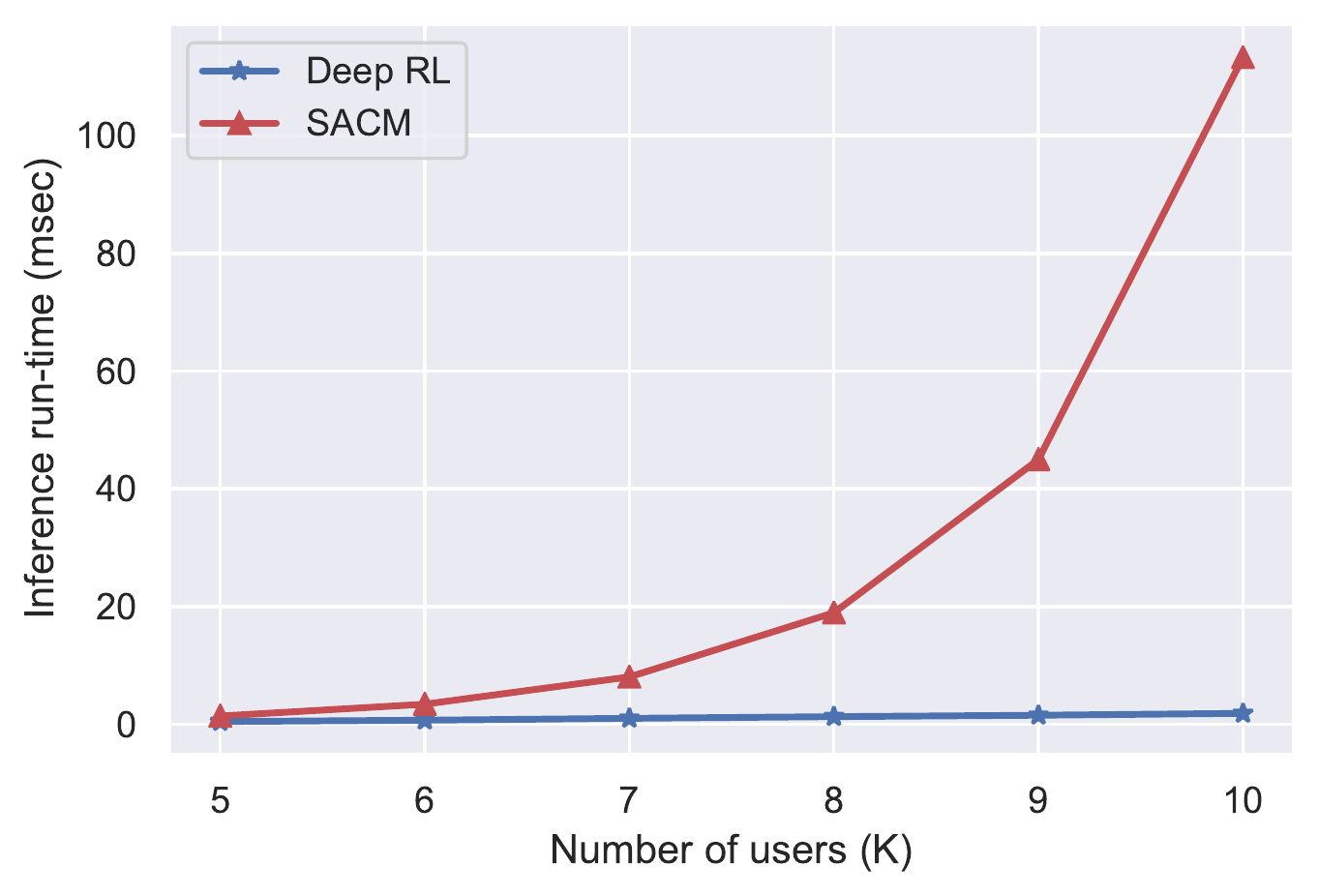}
\caption{Comparison of the inference run-times of the proposed approach and the SACM algorithm on a typical laptop for systems with $5$ to $10$ users, and per-user cache size of $M=4$.}
\label{fig:complexity}
\end{figure}

Figure \ref{fig:complexity} shows an example run-time comparison between our proposed approach and the SACM approach for networks with $5$ to $10$ users, each with a cache of $M=3$ files per user.
As the figure demonstrates, our approach provides a significant reduction in computational complexity compared to the exponential complexity of the SACM approach, while achieving an identical or better performance in terms of normalized delivery delay.

\section{Concluding Remarks}\label{sec:conc}
In this paper, we considered the problem of minimizing the delivery delay in a cache-aided network with a single server and multiple users, each of which has arbitrarily cached parts of the contents from a central library of files. We introduced a method based on deep reinforcement learning, in which an agent at the server interacts with the system by receiving observations on the users' caches and requests, taking actions by forming coded bits to broadcast to the users, and receiving rewards based on how many bits were successfully delivered to the users across the network. We demonstrated the superiority of our proposed approach compared to the state-of-the-art baselines, both in terms of delivery delay and computational complexity.

There are multiple directions following this work. Training the agent in a system with a large number of users is complex and requires a considerable amount of time. Therefore, an interesting future direction would be to leverage the knowledge of the underlying problem and possible useful combinations of bits to make the training process more efficient. Moreover, extending this framework to other domains, such as D2D networks and interference channels, is another exciting direction to pursue, as in those cases, the problem will turn to a multi-agent reinforcement learning problem with its own unique set of challenges.

\balance
\bibliographystyle{IEEEtran}
\bibliography{navid}

\begin{thebibliography}{10}
\providecommand{\url}[1]{#1}
\csname url@samestyle\endcsname
\providecommand{\newblock}{\relax}
\providecommand{\bibinfo}[2]{#2}
\providecommand{\BIBentrySTDinterwordspacing}{\spaceskip=0pt\relax}
\providecommand{\BIBentryALTinterwordstretchfactor}{4}
\providecommand{\BIBentryALTinterwordspacing}{\spaceskip=\fontdimen2\font plus
\BIBentryALTinterwordstretchfactor\fontdimen3\font minus
  \fontdimen4\font\relax}
\providecommand{\BIBforeignlanguage}[2]{{%
\expandafter\ifx\csname l@#1\endcsname\relax
\typeout{** WARNING: IEEEtran.bst: No hyphenation pattern has been}%
\typeout{** loaded for the language `#1'. Using the pattern for}%
\typeout{** the default language instead.}%
\else
\language=\csname l@#1\endcsname
\fi
#2}}
\providecommand{\BIBdecl}{\relax}
\BIBdecl

\bibitem{cisco}
``Cisco visual networking index: Global mobile data traffic forecast update,
  2017-2022 white paper,'' February 2019.

\bibitem{maddah2014fundamental}
M.~A. Maddah-Ali and U.~Niesen, ``Fundamental limits of caching,'' \emph{IEEE
  Transactions on Information Theory}, vol.~60, no.~5, pp. 2856--2867, 2014.

\bibitem{decent}
------, ``Decentralized coded caching attains order-optimal memory-rate
  tradeoff,'' \emph{IEEE/ACM Transactions on Networking (TON)}, vol.~23, no.~4,
  pp. 1029--1040, 2015.

\bibitem{niesen2016coded}
U.~Niesen and M.~A. Maddah-Ali, ``Coded caching with nonuniform demands,''
  \emph{IEEE Transactions on Information Theory}, vol.~63, no.~2, pp.
  1146--1158, 2016.

\bibitem{pedarsani2016online}
R.~Pedarsani, M.~A. Maddah-Ali, and U.~Niesen, ``Online coded caching,''
  \emph{IEEE/ACM Transactions on Networking (TON)}, vol.~24, no.~2, pp.
  836--845, 2016.

\bibitem{karamchandani2016hierarchical}
N.~Karamchandani, U.~Niesen, M.~A. Maddah-Ali, and S.~N. Diggavi,
  ``Hierarchical coded caching,'' \emph{IEEE Transactions on Information
  Theory}, vol.~62, no.~6, pp. 3212--3229, 2016.

\bibitem{asghari}
S.~M. {Asghari}, Y.~{Ouyang}, A.~{Nayyar}, and A.~S. {Avestimehr}, ``An
  approximation algorithm for optimal clique cover delivery in coded caching,''
  \emph{IEEE Transactions on Communications}, vol.~67, no.~7, pp. 4683--4695,
  July 2019.

\bibitem{maddah2016coding}
M.~A. Maddah-Ali and U.~Niesen, ``Coding for caching: fundamental limits and
  practical challenges,'' \emph{IEEE Communications Magazine}, vol.~54, no.~8,
  pp. 23--29, 2016.

\bibitem{niesen2015coded}
U.~Niesen and M.~A. Maddah-Ali, ``Coded caching for delay-sensitive content,''
  in \emph{2015 IEEE International Conference on Communications (ICC)}.\hskip
  1em plus 0.5em minus 0.4em\relax IEEE, 2015, pp. 5559--5564.

\bibitem{yu2017exact}
Q.~Yu, M.~A. Maddah-Ali, and A.~S. Avestimehr, ``The exact rate-memory tradeoff
  for caching with uncoded prefetching,'' \emph{IEEE Transactions on
  Information Theory}, vol.~64, no.~2, pp. 1281--1296, 2017.

\bibitem{wan2016optimality}
K.~Wan, D.~Tuninetti, and P.~Piantanida, ``On the optimality of uncoded cache
  placement,'' in \emph{2016 IEEE Information Theory Workshop (ITW)}.\hskip 1em
  plus 0.5em minus 0.4em\relax IEEE, 2016, pp. 161--165.

\bibitem{shariatpanahi2016multi}
S.~P. Shariatpanahi, S.~A. Motahari, and B.~H. Khalaj, ``Multi-server coded
  caching,'' \emph{IEEE Transactions on Information Theory}, vol.~62, no.~12,
  pp. 7253--7271, 2016.

\bibitem{shariatpanahi2017multi}
S.~P. Shariatpanahi, G.~Caire, and B.~H. Khalaj, ``Multi-antenna coded
  caching,'' in \emph{2017 IEEE International Symposium on Information Theory
  (ISIT)}.\hskip 1em plus 0.5em minus 0.4em\relax IEEE, 2017, pp. 2113--2117.

\bibitem{ji2015wireless}
M.~Ji, G.~Caire, and A.~F. Molisch, ``Wireless device-to-device caching
  networks: Basic principles and system performance,'' \emph{IEEE Journal on
  Selected Areas in Communications}, vol.~34, no.~1, pp. 176--189, 2015.

\bibitem{ji2015fundamental}
------, ``Fundamental limits of caching in wireless {D2D} networks,''
  \emph{IEEE Transactions on Information Theory}, vol.~62, no.~2, pp. 849--869,
  2015.

\bibitem{wan2018caching}
K.~Wan, M.~Ji, P.~Piantanida, and D.~Tuninetti, ``Caching in combination
  networks: Novel multicast message generation and delivery by leveraging the
  network topology,'' in \emph{2018 IEEE International Conference on
  Communications (ICC)}.\hskip 1em plus 0.5em minus 0.4em\relax IEEE, 2018, pp.
  1--6.

\bibitem{maddah2015cache}
M.~A. Maddah-Ali and U.~Niesen, ``Cache-aided interference channels,'' in
  \emph{2015 IEEE International Symposium on Information Theory (ISIT)}.\hskip
  1em plus 0.5em minus 0.4em\relax IEEE, 2015, pp. 809--813.

\bibitem{naderializadeh2017fundamental}
N.~Naderializadeh, M.~A. Maddah-Ali, and A.~S. Avestimehr, ``Fundamental limits
  of cache-aided interference management,'' \emph{IEEE Transactions on
  Information Theory}, vol.~63, no.~5, pp. 3092--3107, 2017.

\bibitem{hachem2018degrees}
J.~Hachem, U.~Niesen, and S.~N. Diggavi, ``Degrees of freedom of cache-aided
  wireless interference networks,'' \emph{IEEE Transactions on Information
  Theory}, vol.~64, no.~7, pp. 5359--5380, 2018.

\bibitem{xu2017fundamental}
F.~Xu, M.~Tao, and K.~Liu, ``Fundamental tradeoff between storage and latency
  in cache-aided wireless interference networks,'' \emph{IEEE Transactions on
  Information Theory}, vol.~63, no.~11, pp. 7464--7491, 2017.

\bibitem{naderializadeh2019cache}
N.~Naderializadeh, M.~A. Maddah-Ali, and A.~S. Avestimehr, ``Cache-aided
  interference management in wireless cellular networks,'' \emph{IEEE
  Transactions on Communications}, vol.~67, no.~5, pp. 3376--3387, 2019.

\bibitem{chen2018caching}
B.~Chen and C.~Yang, ``Caching policy for cache-enabled {D2D} communications by
  learning user preference,'' \emph{IEEE Transactions on Communications},
  vol.~66, no.~12, pp. 6586--6601, 2018.

\bibitem{ndikumana2018deep}
A.~Ndikumana, N.~H. Tran, and C.~S. Hong, ``Deep learning based caching for
  self-driving car in multi-access edge computing,'' \emph{arXiv preprint
  arXiv:1810.01548}, 2018.

\bibitem{sadeghi2019reinforcement}
A.~Sadeghi, F.~Sheikholeslami, A.~G. Marques, and G.~B. Giannakis,
  ``Reinforcement learning for adaptive caching with dynamic storage pricing,''
  \emph{IEEE Journal on Selected Areas in Communications}, vol.~37, no.~10, pp.
  2267--2281, 2019.

\bibitem{zhong2018deep}
C.~Zhong, M.~C. Gursoy, and S.~Velipasalar, ``A deep reinforcement
  learning-based framework for content caching,'' in \emph{2018 52nd Annual
  Conference on Information Sciences and Systems (CISS)}.\hskip 1em plus 0.5em
  minus 0.4em\relax IEEE, 2018, pp. 1--6.

\bibitem{somuyiwa2018reinforcement}
S.~O. Somuyiwa, A.~Gy{\"o}rgy, and D.~G{\"u}nd{\"u}z, ``A
  reinforcement-learning approach to proactive caching in wireless networks,''
  \emph{IEEE Journal on Selected Areas in Communications}, vol.~36, no.~6, pp.
  1331--1344, 2018.

\bibitem{jiang2019multi}
W.~Jiang, G.~Feng, S.~Qin, T.~S.~P. Yum, and G.~Cao, ``Multi-agent
  reinforcement learning for efficient content caching in mobile {D2D}
  networks,'' \emph{IEEE Transactions on Wireless Communications}, vol.~18,
  no.~3, pp. 1610--1622, 2019.

\bibitem{bhatnagar2009natural}
S.~Bhatnagar, R.~S. Sutton, M.~Ghavamzadeh, and M.~Lee, ``Natural actor--critic
  algorithms,'' \emph{Automatica}, vol.~45, no.~11, pp. 2471--2482, 2009.

\bibitem{degris2012model}
T.~Degris, P.~M. Pilarski, and R.~S. Sutton, ``Model-free reinforcement
  learning with continuous action in practice,'' in \emph{2012 American Control
  Conference (ACC)}.\hskip 1em plus 0.5em minus 0.4em\relax IEEE, 2012, pp.
  2177--2182.

\bibitem{williams1991function}
R.~J. Williams and J.~Peng, ``Function optimization using connectionist
  reinforcement learning algorithms,'' \emph{Connection Science}, vol.~3,
  no.~3, pp. 241--268, 1991.

\bibitem{mnih2016asynchronous}
V.~Mnih, A.~P. Badia, M.~Mirza, A.~Graves, T.~Lillicrap, T.~Harley, D.~Silver,
  and K.~Kavukcuoglu, ``Asynchronous methods for deep reinforcement learning,''
  in \emph{International conference on machine learning}, 2016, pp. 1928--1937.

\bibitem{de2016complexity}
J.~De~Cock, Z.~Li, M.~Manohara, and A.~Aaron, ``Complexity-based
  consistent-quality encoding in the cloud,'' in \emph{2016 IEEE International
  Conference on Image Processing (ICIP)}.\hskip 1em plus 0.5em minus
  0.4em\relax IEEE, 2016, pp. 1484--1488.

\end{thebibliography}

\end{document}